\documentclass{aa}
\usepackage{txfonts}
\usepackage{natbib}
\bibpunct{(}{)}{;}{a}{}{,}
\usepackage{graphicx}

\newcommand\jcd{Christensen-Dalsgaard}
\newcommand\ea{et al.}
\newcommand{\hrd}{HR diagram}
\newcommand{\cdd}{\mbox{C-D} diagram}
\newcommand{\dtd}{$\Delta$--$t$ diagram}
\newcommand{\deltt}{$\Delta$--$t$}
\newcommand{\bvf}{Brunt-V\"ais\"al\"a frequency}
\newcommand{\msun}{\ensuremath{M_\odot}}
\newcommand{\rsun}{\ensuremath{R_\odot}}

\newcommand{\rhob}{\ensuremath{\bar{\rho}}}
\newcommand{\rhobs}{\ensuremath{\bar{\rho}_\odot}}
\newcommand{\xc}{\ensuremath{X_\mathrm{c}}}
\newcommand{\dov}{\ensuremath{d_\mathrm{ov}}}
\newcommand{\febyh}{\ensuremath{\mathrm{[Fe/H]}}}
\newcommand{\teff}{\ensuremath{T_\mathrm{eff}}}

\newcommand{\dzz}{\ensuremath{\langle \Delta\nu_0\rangle}}

\newcommand{\dzt}{\ensuremath{\langle d_{02}\rangle}}

\newcommand{\ddzo}{\ensuremath{\langle dd_{01}\rangle}}

\newcommand{\rzt}{\ensuremath{\langle r_{02}\rangle}}
\newcommand{\rzo}{\ensuremath{\langle r_{01}\rangle}}

\newcommand{\tauhiz}{\ensuremath{\tau_\mathrm{HIZ}}}
\newcommand{\taubcz}{\ensuremath{\tau_\mathrm{BCZ}}}
\newcommand{\ttauhiz}{\ensuremath{\tilde{\tau}_\mathrm{HIZ}}}
\newcommand{\ttaubcz}{\ensuremath{\tilde{\tau}_\mathrm{BCZ}}}
\newcommand{\thiz}{\ensuremath{t_\mathrm{HIZ}}}
\newcommand{\tbcz}{\ensuremath{t_\mathrm{BCZ}}}
\newcommand{\tthiz}{\ensuremath{\tilde{t}_\mathrm{HIZ}}}
\newcommand{\ttbcz}{\ensuremath{\tilde{t}_\mathrm{BCZ}}}
\newcommand{\tz}{\ensuremath{t_0}}
\newcommand{\ttz}{\ensuremath{\tilde{t}_0}}

\def\myfigure#1#2#3#4{
	\begin{figure#4}
	\resizebox{\hsize}{!}{\includegraphics{#1}}
	\caption{#2 \label{#3}}
	\end{figure#4}
}

\begin{document}

\title{Asteroseismic diagrams for solar-type stars}

\author{Anwesh Mazumdar\inst{1,2}}

\offprints{anwesh.mazumdar@ster.kuleuven.ac.be}

\institute{Instituut voor Sterrenkunde, K.\ U.\ Leuven, Celestijnenlaan
200B, 3001 Leuven, Belgium
\and
Observatoire de Paris, LESIA, CNRS UMR 8109, 92195 Meudon, France
}

\date{}

\abstract{
We explore the feasibility of applying the \jcd\ diagram to real
asteroseismic data and provide quantitative measures of the uncertainty
associated with the results. We also propose a new kind of seismic
diagram, based on the determination of the locations of sharp acoustic
features inside a star. We show that by combining the information about
the position of the base of the convective envelope or the \ion{He}{ii}
ionisation zone with a measure of the average large separation, it is
possible to constrain the unknown chemical composition or the various
parameters characterising the physical processes in the stellar
interior. We demonstrate the application of this technique to the
analysis of mock data for a CoRoT target star.

\keywords{stars: oscillations -- stars: interiors}
}

\maketitle

\section{Introduction}
\label{sec:intro}

Seismic observations of distant stars, either from the ground or space,
is likely to permit the detection of only low degree modes of
oscillations ($\ell =0,1,2,3$).  In this context, it is necessary to
develop or adapt the existing seismic tools of analysis to extract the
maximum amount of information contained in these modes. One approach is
through the construction of suitable asteroseismic diagrams -- the
technique of combining useful features of the frequency spectrum of
stars to extract information about the physical processes in the
stellar interior that affect the frequencies. The basic idea behind
such diagrams is that they are to be constructed in terms of
asteroseismic observables and must reveal some structural or
evolutionary feature of a star through a calibration of theoretical
stellar models.

The most well-known asteroseismic diagram is the so-called \jcd\
diagram, or the \cdd, in short \citep{jcd88}.  This diagram exploits
the fact that while the average large separation of radial modes, $\dzz
\equiv \langle \Delta\nu_0(n) \rangle = \langle \nu(n,0) - \nu(n-1,0)
\rangle $ reflects the gross properties of a star, like its mass and
radius, the average small separation, $\dzt \equiv \langle d_{02}(n)
\rangle = \langle \nu(n,0) - \nu(n-1,2)\rangle$ is more sensitive to
the innermost layers of the star, and therefore to its evolutionary
state.  The averages are computed over several high-order modes (the
so-called asymptotic range).  By determining the position of a star on
this diagram, through measurement of the large and small separations of
its oscillation frequencies, it is possible to determine its mass and
age.

In this work, we study the feasibility of using the \cdd\ for real
asteroseismic data containing finite errors in the frequencies. We
investigate the accuracy to which the mass and age of a star with
unknown characteristics (such as the chemical composition or the extent
of convective overshoot) can be predicted from the \cdd\
(Section~\ref{sec:cdd_param}). 

The rest of the paper introduces a new kind of asteroseismic diagram.
The acoustic location of sharp features inside a star such as the
boundaries of convective zones or ionisation zones can be determined
using the oscillatory signal in the frequencies that they produce
\citep[e.g.,][]{mct94,ban94,rv94}.  Considering that the location of
these regions cannot be independent of the general stratification
inside a star, in Section~\ref{sec:dtd} we propose a new diagram,
connecting these independent seismic observables and the mean large
separation.  An application of this seismic diagram to the analysis of
synthetic data for the star \object{HD49933} illustrates how it could
help substantially in constructing a reliable model of the star
(Section~\ref{sec:hd49933}).

This work utilises several grids of stellar models with different
stellar parameters which were constructed using the CESAM evolutionary
code \citep{morel97}. These models used the OPAL equation of state
\citep{rn02} and OPAL opacity tables \citep{ir96}, complemented by the
low-temperature opacity tables of \citet{af94}. Convection was
described by the standard mixing length theory \citep{hvb65} and
nuclear reaction rates were obtained from the NACRE compilation
\citep{aar99}.  Diffusion of helium and heavy elements was not
considered for the present study.  While it is generally accepted that
some form of diffusive phenomenon is very likely to exist in stellar
interiors, we believe that its inclusion would not alter the results of
this work in a qualitative way. The effect of incorporating diffusion
would be akin to exploring another dimension in stellar parameter space
-- the various calibration curves might change in absolute terms, but
that would not affect the general conclusions drawn from them. 

The p-mode oscillation frequencies of low degree modes ($\ell=0,1,2,3$)
were computed for each model under the adiabatic approximation, using
the Aarhus pulsation package, ADIPLS \citep{cb91}. Throughout the work,
we have used these low degree modes which are the only ones relevant
for real asteroseismic data.

\section{Sensitivity of the \cdd\ to stellar parameters}
\label{sec:cdd_param}

The general idea of the \cdd\ has already been presented in
Section~\ref{sec:intro}.  Recently, \citet{rv03} have demonstrated that
while the frequency separations themselves are affected by the effects
of the surface layers, the ratio of the small separation to the large
separation, defined as $\rzt \equiv \langle r_{02}(n) \rangle = \langle
\frac{d_{02}(n)}{\Delta\nu_1(n)} \rangle$, can exclude such effects to
a great extent.  Similar \cdd s can be constructed using these ratios,
instead of the small separations themselves \citep{fct05}.  A
comparison of the classical \cdd\ and its variation for $\ell=0,1$
modes \citep{mr03}, with small separations as well as the ratio of
separations, can be found in the online appendix.

To investigate whether the \cdd\ can indeed be used for real
asteroseismic data, we compare the sensitivity of the tracks on the
\cdd\ toward different stellar parameters to the uncertainty associated
with placing an observed star on this diagram. We estimate the errors
in \dzz\ and \dzt\ through a Monte Carlo simulation where random errors
with a specified standard deviation ($\sigma_\nu = 10^{-4}\nu$) are
added to the exact frequencies before computing the separations. Such
an error margin in the frequencies is consistent with the expectations
from the current and future asteroseismic space missions, MOST
\citep{walker03} and CoRoT \citep{baglin03}, as well as recent
ground-based seismic observations \citep[e.g.,][]{bk03}.  The variance
in the average values of the large and small separations are estimated
to be the uncertainties in these quantities for a real data set with
comparable errors in frequencies.  We find that the error in
determining the average large separation is only $\sim 0.02~\mu$Hz.
Thus, although the individual frequency values might have errors up to
$\sim 0.3~\mu$Hz, the average value, \dzz\ over a suitably chosen range
of radial orders is quite a robust quantity.  On the other hand, the
uncertainty in the average small separation, \dzt\ is $\sim
0.0015~\mu$Hz. Very similar estimates for these uncertainties were
found by \citet{apc95}.  However, the errors in determining these
quantities would be affected by the total number of modes observed as
well as the mass and age of the star. At late stages of evolution,
particularly in more massive stars, the presence of mixed non-radial
modes reduces the number of modes available for such asymptotic
analysis. Typically, this causes an increase in the error on the
average small separations. The error estimates quoted here, and shown
in Figs.~\ref{fig:dr02_alpha} to~\ref{fig:dr02_chem} are, therefore,
only representative.  

The theoretical \cdd\ is typically based on stellar models constructed
with a particular choice of parameters, principal among which are the
mixing length $\alpha$, the extent of convective core overshoot, \dov\
(both $\alpha$ and \dov\ being measured in terms of local pressure
scale height) and the initial chemical composition (characterised by
($X_0,Z_0$)).  In order to investigate the effect of these parameters
on the \cdd, we constructed separate grids of stellar models with
different values for these parameters.  For each of these sets of
models, only one parameter (e.g., $\alpha$) is varied at a time,
keeping the others (viz., \dov, $X_0$, $Z_0$) identical, so that each
effect can be separately understood.  The frequency range for
determining the asymptotic average values of the separations is chosen
in terms of the scaled frequency, $\nu/\sqrt{(M/\msun)/(R/\rsun)^3}$,
where $M$ and $R$ are the mass and radius for each model respectively,
to be 1.5~mHz to 3.5~mHz. This typically corresponds to averaging over
15 radial orders.

In Fig.~\ref{fig:dr02_alpha}, we show the variation in the \cdd s of
\dzt\ and \rzt\ with the mixing length parameter, $\alpha$. Each curve
on this diagram (and Figs.~\ref{fig:dr02_ov} and \ref{fig:dr02_chem})
is an evolutionary track of a given mass, indicated in solar units at
the ZAMS end of the track. We consider two sets of models,
corresponding to $\alpha = 1.8$ and 1.6.  Since the outer convective
envelope grows thinner for higher masses, naturally, the effect of
changing the mixing length reduces with mass.  For low mass stars, the
effect is comparable to the errors in \dzt\ or \rzt, but the tracks are
still close enough to allow an estimate of the stellar mass within
$5\%$. 
\myfigure{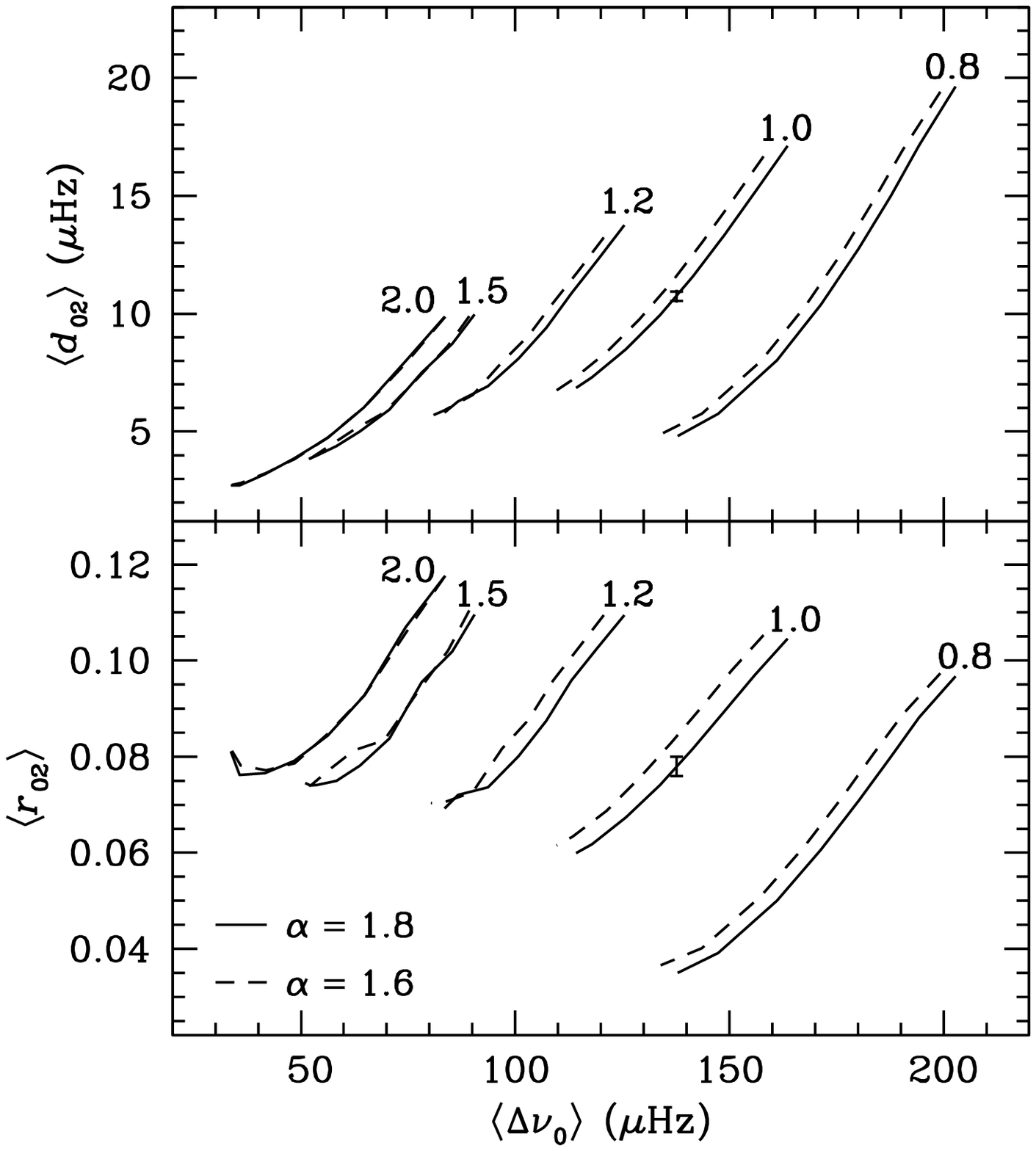}{The variation of the \cdd\ using the average
small separations, \dzt\ ({\it top panel}), and the average ratios,
\rzt\ ({\it bottom panel}) with the choice of the mixing length
parameter, $\alpha$, is shown. Two sets of models have been used -- one
with $\alpha = 1.8$ ({\it solid curves}) and another with $\alpha =
1.6$ ({\it dashed curves}) to illustrate the shift in the tracks. All
other stellar parameters are identical in the two sets of models.
Representative errorbars corresponding to relative frequency errors of
$10^{-4}$ are shown.}{fig:dr02_alpha}{}

The effect of convective core overshoot on the \cdd\ is more dramatic.
We show this effect in Fig.~\ref{fig:dr02_ov} by comparing models
without overshoot ($\dov = 0$) and those with $\dov = 0.2$.  The effect
of overshoot in increasing the total radius of the star at a given
evolutionary stage (in terms of the central hydrogen abundance, \xc) is
clearly seen in the extension of the tracks on the \cdd\ to lower
values of \dzz.  Further, the value of \dzt\ also reduces with
overshoot, reflecting the higher age of the star. In effect, the tracks
on the \cdd\ deviate considerably from the standard ones when
overshooting is included in the models. It is evident, therefore, that
we need careful calibration of the effect of overshoot if one is to
estimate the mass and age of a star (with mass $\ga 1.1\msun$) directly
from the \cdd. 
\myfigure{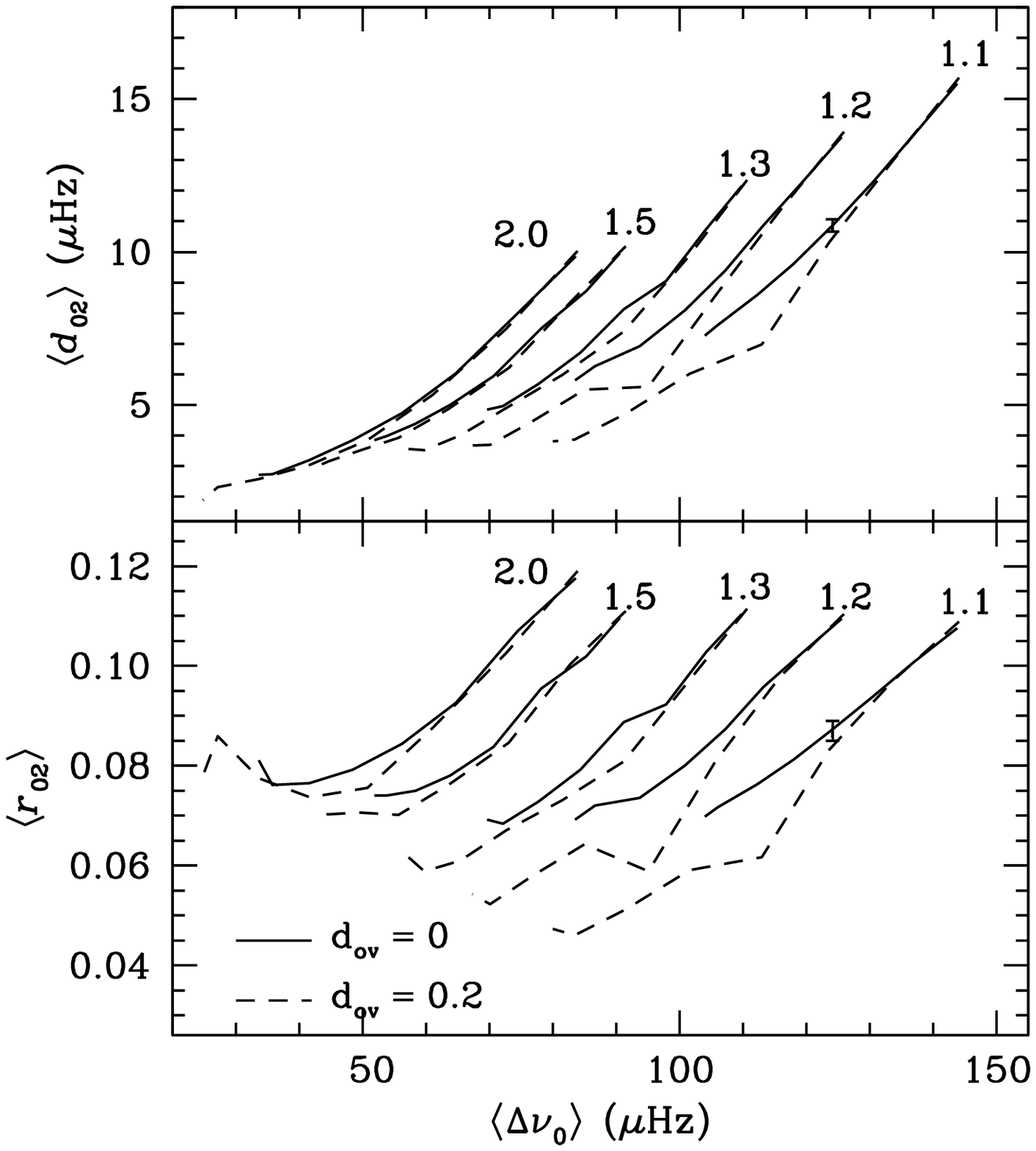}{The variation of the \cdd\ using the average
small separations, \dzt\ ({\it top panel}), and the average ratios,
\rzt\ ({\it bottom panel}) with convective core overshoot is shown. Two
sets of models have been used -- one with $\dov = 0$ ({\it solid
curves}) and another with $\dov = 0.2$ ({\it dashed curves}) to
illustrate the shift in the tracks. All other stellar parameters are
identical in the two sets of models. Representative errorbars
corresponding to relative frequency errors of $10^{-4}$ are
shown.}{fig:dr02_ov}{}

While the metallicity of a target star is often known from
spectroscopic data, its helium content is typically unknown, thereby
one is left with the ambiguity of one unknown parameter concerning the
chemical composition.  We consider three different scenarios which
approximately reflect the various possibilities for the chemical
composition of a star with a known metallicity (or \febyh).  We compare
the tracks on the \cdd\ for all these cases with respect to those for
the standard solar composition in Fig.~\ref{fig:dr02_chem}.

The first case ($(X_0,Z_0) = (0.73,0.01)$) corresponds to a typical
metal-poor star ($\febyh \sim -0.25$) with proportionally reduced
helium content. This would exemplify the situation where we would only
have an indirect estimate of the helium abundance from the metallicity,
based on assumptions about the chemical evolution of the galaxy.  The
second combination ($(X_0,Z_0) = (0.70,0.03)$) mimics a star which is
metal-rich ($\febyh \sim 0.25$), but has near-solar helium abundance.
The third case ($(X_0,Z_0) = (0.73,0.02)$) reflects the chemical
composition of a star which has solar metallicity, but is known to be
over-abundant in hydrogen.
\myfigure{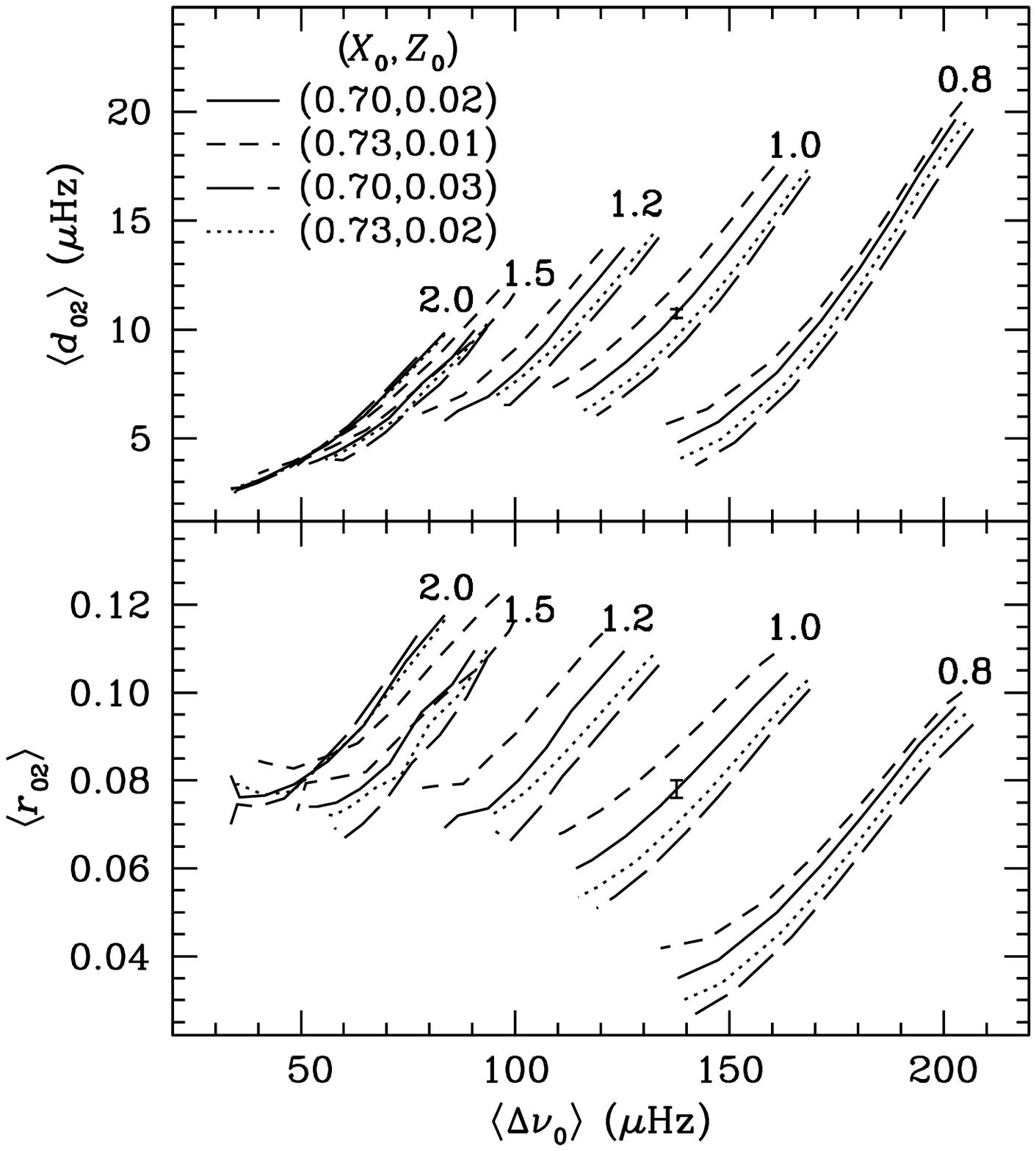}{The variation of the \cdd\ using the average
small separations, \dzt\ ({\it top panel}), and the average ratios,
\rzt\ ({\it  bottom panel}) with initial chemical composition is shown.
Four sets of models with different combinations of $(X_0,Z_0)$ have
been used -- $(0.70,0.02)$ ({\it solid curves}), $(0.73,0.01)$ ({\it
small dashed curves}), $(0.70,0.03)$ ({\it long dashed curves}), and
$(0.73,0.02)$ ({\it dotted curves}) to illustrate the shift in the
tracks. All other stellar parameters are identical in the two sets of
models. Representative errorbars corresponding to relative frequency
errors of $10^{-4}$ are shown.}{fig:dr02_chem}{}
It is evident that in each case, the tracks depend substantially on the
chemical composition. It appears that even a slight change in the
balance of helium and heavier elements at a constant hydrogen abundance
produces a large shift in the \cdd.  In summary, the effectiveness of
the \cdd\ may be strongly undermined if the chemical composition is not
fully known. 

The importance of the equation of state (EOS) in the determination of
internal stellar structure and its seismic manifestation cannot be
ignored. However, by comparing two sets of models with OPAL and EFF
\citep{eff73} EOS we find that the shift in the tracks are at a level
much lower than that induced by the other parameters discussed above,
especially at higher masses.  Therefore, it would be fair to deduce
that the choice of the EOS would not greatly affect the extraction of
the mass and age of a star from the \cdd.

\section{$\Delta$--$t$ diagram}
\label{sec:dtd}

\subsection{Probing sharp acoustic features}
\label{subsec:acdep}

While the \cdd\ is indeed a powerful tool to estimate the mass and age
of a star from its oscillation frequencies, the process of averaging
the frequency separations actually reduces the information content of
the frequency spectrum. One of the other means to investigate the
structure of the stellar interior from the oscillation frequencies
without resorting to explicit modelling of the star is to utilise the
oscillatory signal in the frequencies to determine the acoustic depth
\begin{equation}
\tau_d = \int^R_{r_d} {dr\over c}\;,
\label{eq:tau}
\end{equation}
of a sharp feature, lying at a radius of $r_d$, $c$ being the sound
speed and $R$ the total radius of the star \citep[e.g.,][]{gough90}.
 
The acoustic depth of the \ion{He}{ii} ionisation zone (hereafter
referred to as HIZ) and the base of the convective envelope (hereafter
referred to as BCZ) can be easily estimated using the second
differences of the frequencies which are defined as
$\delta^2\nu_{n,\ell}=\nu_{n+1,\ell}-2\nu_{n,\ell}+\nu_{n-1,\ell}$\,
\citep{ma01}.  Following \citet{basu04}, we fit a function of the form
\begin{eqnarray}
&&\delta^{2}\nu=\left(a_1+ a_2\nu +{a_3\over\nu^2}\right) \nonumber\\
&&\qquad\qquad +
  \left(b_1+{b_2\over\nu}+{b_3\over\nu^2}\right)
  \sin(4\pi\nu\ttauhiz+\phi_{\rm {HIZ}})\; \nonumber\\
&&\qquad\qquad +
  \left(c_1+{c_2\over\nu}+{c_3\over\nu^2}\right)
  \sin(4\pi\nu\ttaubcz+\phi_{{\rm BCZ}}),
\label{eq:fit}
\end{eqnarray}
to the second differences of the frequencies. The parameters $a_i, b_i,
c_i~(i=1,2,3)$ and $\ttauhiz, \ttaubcz, \phi_{\rm {HIZ}}$ and
$\phi_{\rm {BCZ}}$ are obtained from a least-squares fit. The frequency
range chosen for performing the fit is 1.5--3.0~mHz, after scaling by
the factor $\dzz/\dzz_\odot$, where $\dzz$ and $\dzz_\odot~(\simeq
135\mu \mathrm{Hz})$ are the average large separations for the
concerned star and the Sun respectively. 

However, it is generally found that the acoustic depths of the BCZ and
HIZ derived from the oscillatory signal in the frequencies (\ttaubcz\
and \ttauhiz) are systematically higher than their true values
(\taubcz\ and \tauhiz) calculated directly from the model according to
the definition, Eq.~(\ref{eq:tau}). This is due to the fact that the
estimation of the acoustic depth of the layer of discontinuity through
the oscillatory signal in frequencies is biased by the surface phase
shift \citep{vbp91}.  However, as pointed out by \citet{cmt95} and
\citet{btg04}, this bias can be largely eliminated in the estimation of
the acoustic radii, \ttbcz\ and \tthiz\ of the BCZ and HIZ, instead of
the acoustic depths:
\begin{equation}
\ttbcz = \ttz - \ttaubcz\;,~\mathrm{and}~~~\tthiz = \ttz - \ttauhiz\;,
\label{eq:ttbcz}
\end{equation}
where $\ttz\ \equiv \frac{1}{2\dzz}$ is an estimate of the total
acoustic radius of the star, $\tz = \int_0^R dr/c$.

By estimating the acoustic radii, \ttbcz\ and \tthiz, for several
models of different mass and age by this method, we find that the true
acoustic radii and the estimated values differ by $\la 5\%$ -- a
remnant effect of the surface phase shift. The differences tend to
increase with mass, because for stars more massive than $\sim 1.5\msun$
the outer convective envelope becomes thin and the BCZ and the HIZ lie
close to each other. In such a situation, it is difficult to fit the
two oscillatory components separately.  Therefore, this method would
not work very well for stars more massive than 1.5\msun.

This difference between the model values of the acoustic radii and
their estimated values from the oscillatory signal exists even for
exact frequencies and might be considered as a systematic bias in the
process. In addition, there would be a certain random error in the case
of actual frequency data associated with observational uncertainty. We
have estimated this error through Monte Carlo simulations similar to
that described in Section~\ref{sec:cdd_param}. We find that the
uncertainty in the estimates of the mean density-scaled acoustic radii
of the BCZ and the HIZ is typically around 40~s for relative frequency
errors of $10^{-4}$. Since the uncertainty in the average large
separation turns out to be tiny (Section~\ref{sec:cdd_param}),
practically the entire contribution towards this error is from the
fitting of \ttaubcz\ and \ttauhiz. This is similar to the error margins
provided by \citet{basu04}, and comparable to that for the BCZ
estimated by \citet{btg04}.

\subsection{A new seismic diagram}
\label{subsec:dtd}

We have seen in Section~\ref{sec:cdd_param} that the average large
separation, which is an index of the gross properties such as the mean
density or the total acoustic size, can be determined to a fairly good
accuracy from an observed frequency spectrum of a solar-type star.
Further, in Section~\ref{subsec:acdep}, we established that the
acoustic position of the BCZ and the HIZ can be determined to an
accuracy of at least $95\%$ from the oscillatory signal in the
frequencies despite the presence of some systematic bias due to unknown
surface effects.  Importantly, these estimates are obtained without
explicit modelling of the stellar structure.  However, the acoustic
location of these layers inside the star cannot be independent of the
general stratification or the chemical composition of the star.
Therefore, the model-independent estimates of \ttbcz\ and \tthiz\
indeed provide clues about the internal structure of the star.
\myfigure{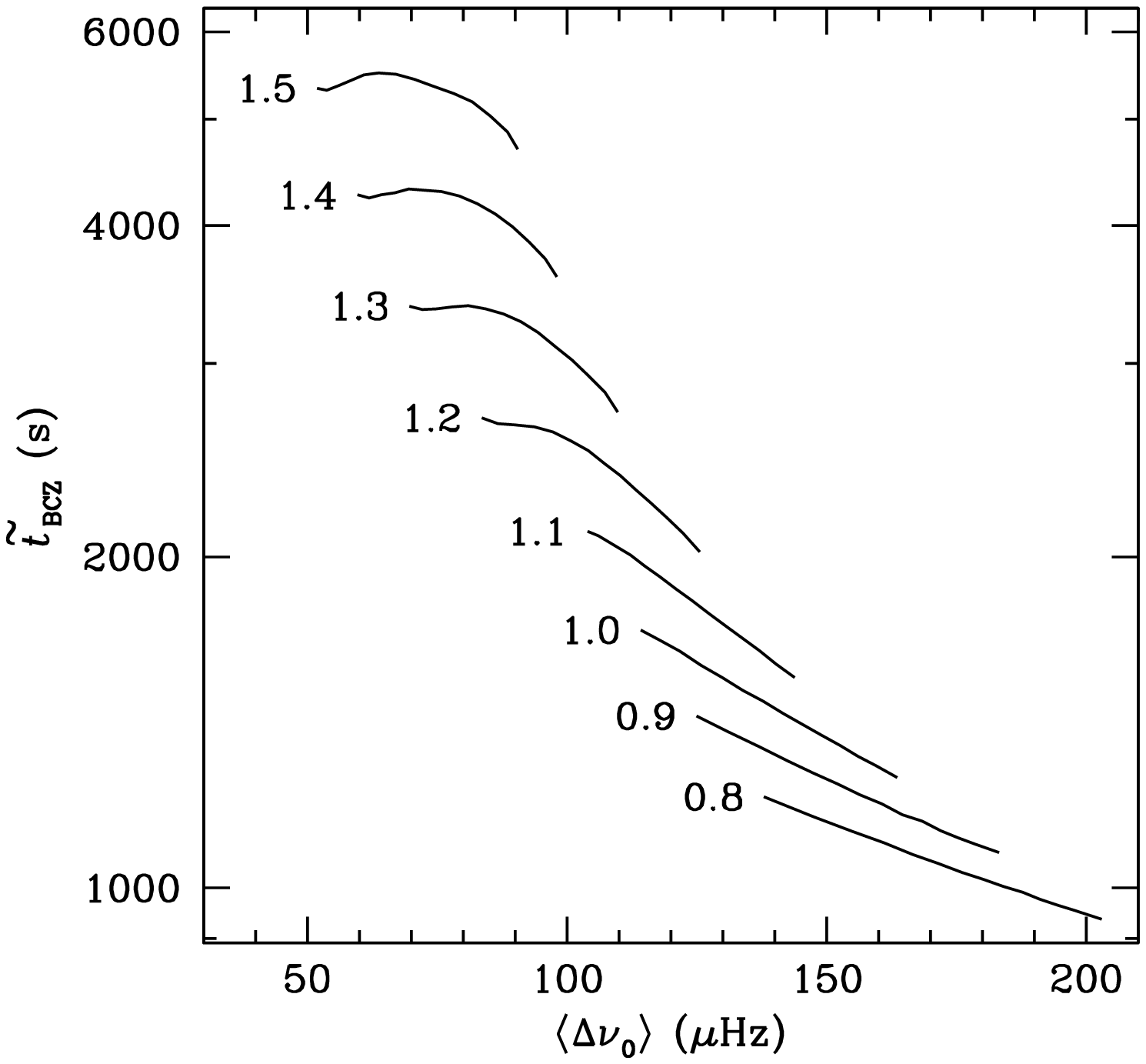}{An example of the \dtd, where the acoustic
radius of the BCZ is plotted as a function of the average large
separation. The ordinate is plotted in logarithmic values for the sake
of clarity. Each curve on this diagram is an evolutionary track of the
indicated mass in solar units.}{fig:delt_mass}{}

In order to combine these independent pieces of information about the
gross properties and the detailed structure of a star, we construct a
diagram connecting the average large separation and the acoustic radius
of the BCZ or the HIZ. The example of such a diagram, which we shall
call the \dtd, is given in Fig.~\ref{fig:delt_mass}. Each line on this
diagram is actually a locus of the acoustic radius of the BCZ (or the
HIZ) as a function of the average large separation as a star of a given
mass evolves on the main sequence. Thus it may be considered as a kind
of seismic diagram, analogous to the \cdd. In this diagram, the
ordinates are actually not the true acoustic radii \tbcz\ and \thiz,
but rather are the values derived from the acoustic depths \taubcz\ and
\tauhiz\ (model values) and the mean large separation \dzz\ (derived
from exact frequencies), through Eq.~(\ref{eq:ttbcz}). This is done
deliberately to minimise the systematic error described in
Section~\ref{subsec:acdep} in the calibration process.  A diagram
similar to that of Fig.~\ref{fig:delt_mass} may be constructed using
the acoustic radii \tthiz, instead of \ttbcz. Since the properties of
these diagrams are, in general, quite similar, it is enough to focus on
the $\Delta$--\ttbcz\ in the following discussion.

\subsection{Sensitivity of \dtd\ to stellar parameters}
\label{subsec:dtd_param}

It is evident from Fig.~\ref{fig:delt_mass} that the \dtd\ has the
diagnostic capability to predict the mass and evolutionary status of a
star, in a manner similar to that of the \cdd.  However, the tracks on
this theoretical diagram would naturally depend on all the parameters
involved in the stellar modelling.  Direct determination of the mass
and age would only be possible if the other stellar parameters are
known {\it a priori}. Since typically this is not the case for real
stars, we have investigated the sensitivity of the \dtd\ to stellar
parameters by using several grids of models, for four different masses,
0.8, 1.0, 1.2 and 1.5\msun.  Since the acoustic radii \ttbcz\ and
\tthiz\ cannot be reliably obtained for more massive stars, they are
not relevant for this exercise. 

In Fig.~\ref{fig:delt_alpha}, we show the \deltt\ tracks for models
with two different values of the mixing length parameter, $\alpha =
1.8$ and 1.6. The errorbars for \ttbcz\ for relative frequency errors
of $10^{-4}$ are magnified by a factor of 3 in these diagrams to
increase visibility. We find that the tracks for \ttbcz\ as a function
of \dzz\ are shifted almost parallelly as $\alpha$ is changed. The
amount of this shift depends on the mass -- while at lower masses it is
too small to be detected given the uncertainty in the seismic
determination of \ttbcz, at masses $\ga 1.2\msun$, the shift is quite
large. This would imply that this diagram is sensitive to $\alpha$ at
higher masses, and might provide at least an indicative value of the
same, provided other parameters are known. 
\myfigure{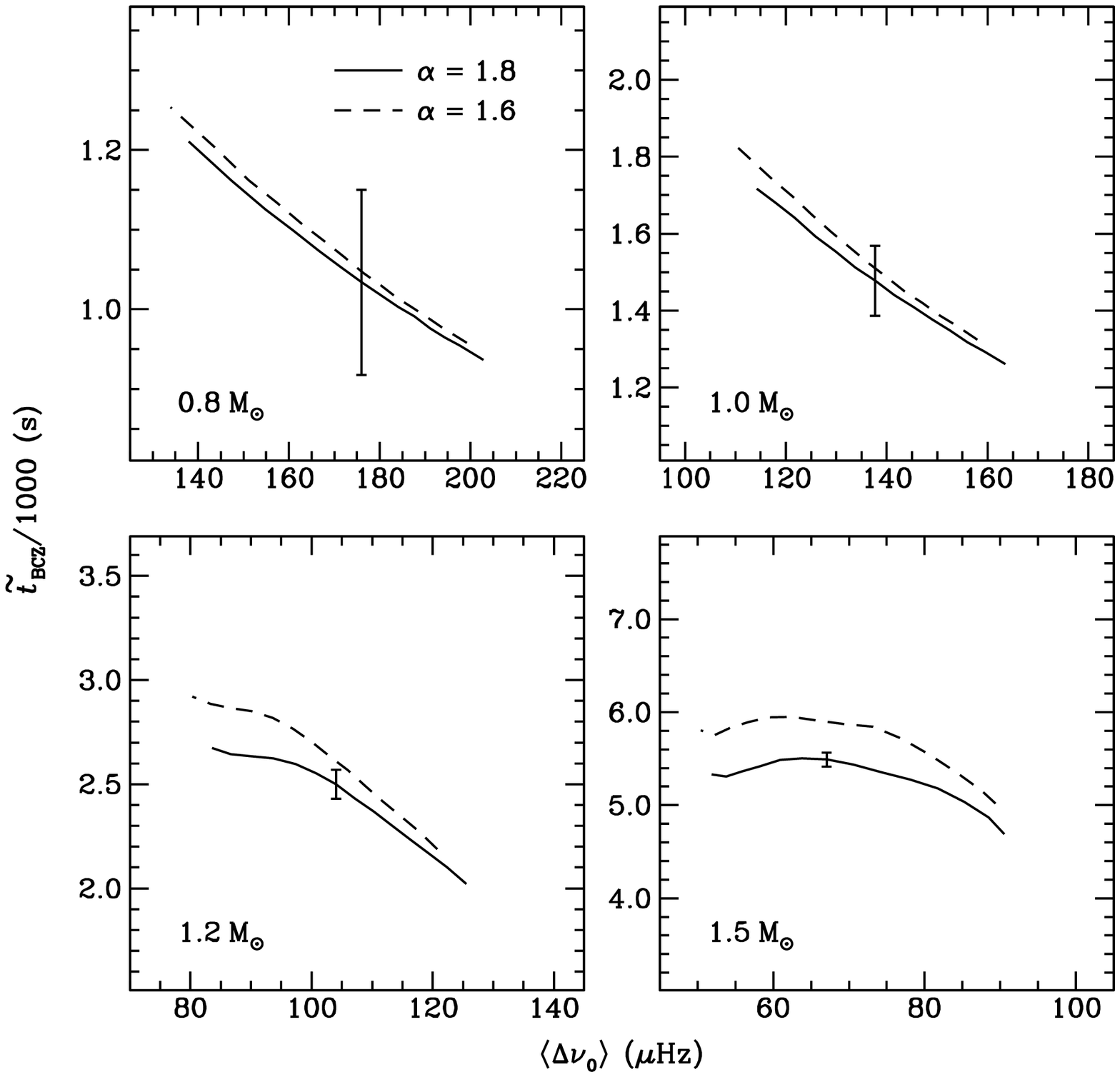}{The sensitivity of the \dtd\ with respect to
the mixing length parameter $\alpha$ is illustrated.  For each mass,
indicated in the different panels, two sets of models have been used --
one with $\alpha = 1.8$ ({\it solid curves}) and another with $\alpha =
1.6$  ({\it dashed curves}).  All other stellar parameters remain
identical.  Representative $3\sigma$ errorbars for the values of the
acoustic radii for a relative error of $10^{-4}$ in frequencies are
shown.}{fig:delt_alpha}{}

The effect of convective core overshoot is expected to be small on
these \dtd s, since the BCZ lies far apart from the convective core.
Figure~\ref{fig:delt_ov}, however, confirms this only partially.  We
find that \ttbcz\ is indeed sensitive to \dov\ at later stages of
evolution for masses between 1.2 and 1.5\msun. 
\myfigure{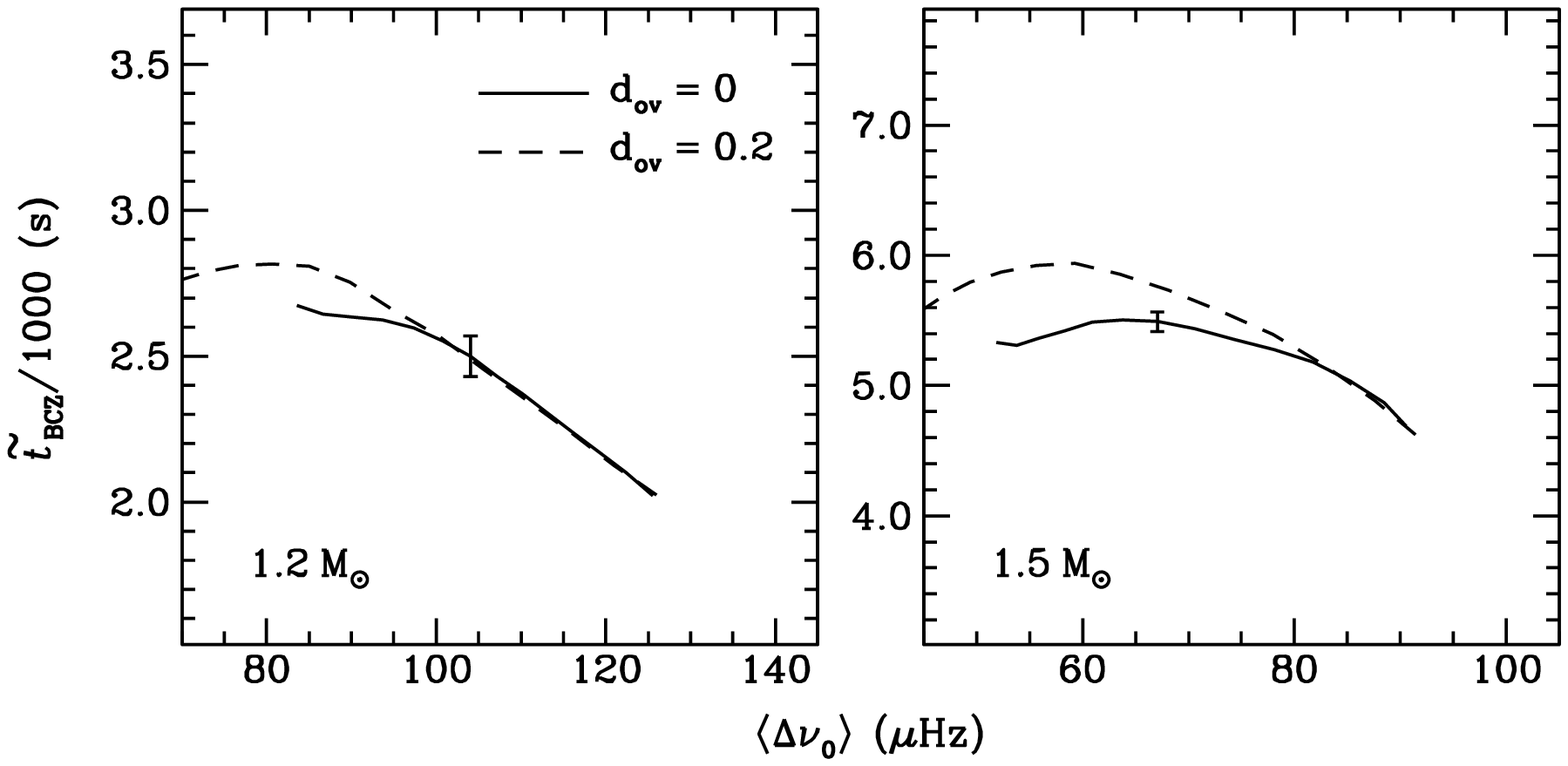}{The sensitivity of the \dtd\ to convective core
overshooting is illustrated. Two sets of models have been used -- one
with $\dov = 0$ ({\it solid curves}) and another with $\dov = 0.2$
({\it dashed curves}).  All other stellar parameters remain identical.
Representative $3\sigma$ errorbars for the values of the acoustic radii
for a relative error of $10^{-4}$ in frequencies are
shown.}{fig:delt_ov}{}

Figure~\ref{fig:delt_chem} illustrates the effects of chemical
composition on the \dtd. The compositions chosen for comparison to the
standard one are identical to the ones described in
Section~\ref{sec:cdd_param}. We find that the BCZ location is sensitive
to the chemical content in all three scenarios. The effect is very
pronounced at higher masses, which would mean that the \dtd\ may be
used to constrain the chemical abundances, at least for 1.2--1.5\msun\
stars.
\myfigure{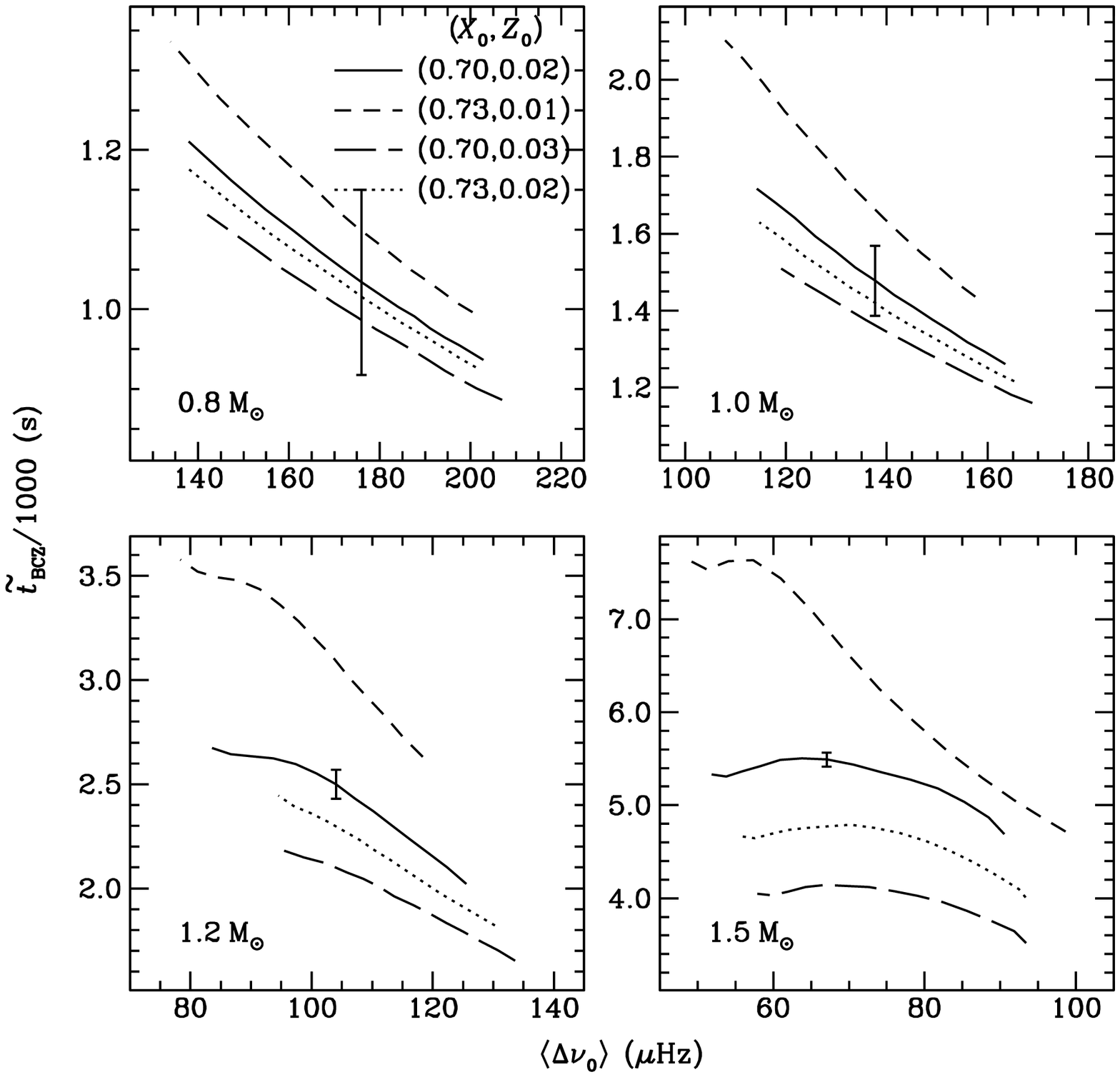}{The sensitivity of the \dtd\ to initial
chemical composition is illustrated. Four sets of models with different
combinations of $(X_0,Z_0)$ have been used -- $(0.70,0.02)$ ({\it solid
curves}), $(0.73,0.01)$ ({\it small dashed curves}), $(0.70,0.03)$
({\it long dashed curves}), and $(0.73,0.02)$ ({\it dotted curves}).
All other stellar parameters remain identical.  Representative
$3\sigma$ errorbars for the values of the acoustic radii for a relative
error of $10^{-4}$ in frequencies are shown.}{fig:delt_chem}{}

\section{Application of the \dtd: The hare \& hound exercise for 
\object{HD49933}}
\label{sec:hd49933}

There might be two approaches to using the \dtd. The first is to
determine any one parameter among $M$, $(X_0,Z_0)$, $\alpha$ or \dov\
directly. But this would require independent knowledge about the
remaining parameters, since each of these parameters affect the \dtd\
considerably. Some constraints about the mass, age and the chemical
composition might be independently gathered in case the target star is
a member of a binary system or a cluster.  In such cases, it might be
possible to use the \dtd\ to estimate the mixing length or the
overshoot parameter.

However, we can also envisage the use of the \dtd\ in another way, when
none of the stellar parameters are known {\it a priori}. In this case,
given an initial trial model, one can test its relative position on the
\dtd\ with respect to the observed data to gain insight into how a
particular parameter needs to be changed from the initial guess in
order to find a better match with the data. This process has to be
iterative, but would considerably reduce the task of searching in a
multi-dimensional parameter space for the best model.

In this section, we illustrate the use of the \dtd\ as applied
successfully in the seismic modelling of a CoRoT primary target star,
\object{HD49933}, based on synthetic data in the course of a hare and
hound exercise. This exercise, carried out within the CoRoT Seismology
Working Group, was conducted in four principal steps --
(i)~construction of a theoretical model for a star, based on its
position in the \hrd\ and available information about its chemical
composition, and calculation of its low-degree eigenfrequencies,
(ii)~conversion of the frequencies into a time series, taking into
account the duration of CoRoT observation (150 days) and the expected
instrumental and background stellar noise, (iii)~extraction and mode
identification of frequencies from this simulated time series, without
knowledge of the original exact frequencies, and (iv)~seismic
interpretation of the extracted frequencies without knowledge of the
original input model. Each of these steps were carried out by
independent groups, maintaining confidentiality of the original input
data, thereby simulating a situation which would be encountered for
real data gathered from the CoRoT mission. The general description of
this exercise can be found on the web
(http://www.ias.u-psud.fr/virgo/html/corot/datagroup/hh.html) and
here we shall restrict ourselves only to the particular case of the
seismic modelling of the star \object{HD49933}\footnote{The relevant
work in steps (i), (ii) and (iii) for \object{HD49933} was carried out
by I.~Roxburgh, C.~Barban and T.~Toutain respectively. The seismic
interpretation in step (iv) described here was carried out by the
author in collaboration with E.~Michel.}. 

The basic parameters available for \object{HD49933} are summarised in
Table~\ref{tab:hd49933} (stage~0). The data on which the seismic
interpretation was based consisted of 67 individual frequencies of
degree $\ell = 0$ and~1, lying between 100 and 3100~$\mu$Hz, extracted
in step (iii) described above. The errors in the frequencies ranged
between 0.02 and 0.46~$\mu$Hz, with the highest errors reported for
intermediate frequencies ($\sim 2000~\mu$Hz).  The average large
separation, \dzz, was found to be $90.4\pm 0.2~\mu$Hz.  Since the mean
density of the star is not known at this stage, the average large
separation was calculated for several different limits in the entire
asymptotic range, and the variation in its value was included as an
additional error. The mean density was estimated to be $0.465\pm
0.015~\rhobs$ through a calibration of the average large separation.
Unfortunately, only a few frequencies for $\ell=2$ and~3 could be
extracted from the simulated time series, and therefore, we had no
small separations $d_{02}$ to utilise in our seismic analysis. This
automatically ruled out the use of the \cdd.

We derived the acoustic depths \ttaubcz\ and \ttauhiz\ by fitting an
expression, similar to that of Eq.~(\ref{eq:fit}) to the second
differences of the frequencies, as shown in Fig.~\ref{fig:hd49933_fit}.
For this exercise, we used all modes between 700 and 2700~$\mu$Hz, the
range being chosen to use as many modes as possible, while not
deviating too much from the standard choice as explained by
\citet{basu04}.  This yielded values of $\ttbcz=4085\pm 68$~s and
$\tthiz=4866\pm 47$~s. The comparison of these results to the original
model values are shown in Table~\ref{tab:hd49933} (stage~1).  The
excellent agreement of these values, obtained prior to any modelling of
the star, indeed confirms the diagnostic power of the oscillatory
signal in the frequencies. In practice, however, the original values
were not revealed at this stage, but only at the end of the entire
exercise.
\myfigure{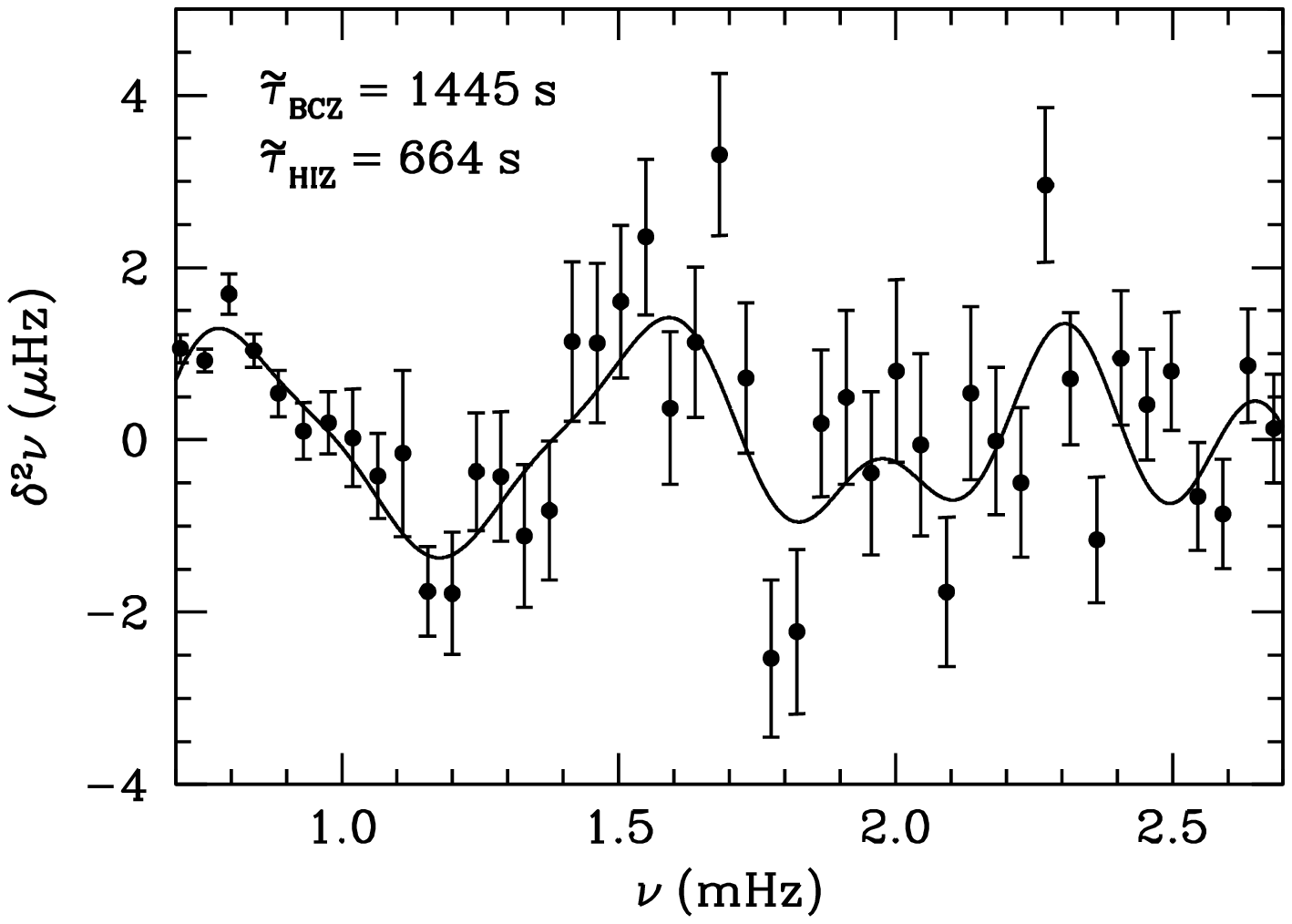}{The oscillatory signal in the second
differences of the frequencies of \object{HD49933} (shown as data
points with respective errorbars) are fitted by the expression
(\ref{eq:fit}) ({\it solid curve}) to extract the values of \ttaubcz\
and \ttauhiz.}{fig:hd49933_fit}{}

In the next step, we proceeded to use the \dtd s to arrive at the
closest model for the star, already taking into account the constraints
imposed on the stellar parameters by the position of \object{HD49933}
on the \hrd. In the top panel of Fig.~\ref{fig:hd49933_hrdt}, we show
the evolutionary tracks of three sets of models with the same mass,
overshoot and the mixing length parameter, but differing in chemical
composition, and passing through the errorbox of \object{HD49933}. The
\dtd\ is illustrated in the bottom panel, where {\em the range in the
values of \dzz\ and \ttbcz\ for each track are restricted to correspond
to their overlap with the box on the \hrd}.  The box on the \dtd\
represents the values of \dzz\ and \ttbcz\ that we have already derived
from the frequencies. This is an example of one of the \dtd s that we
utilised in this analysis, in each case varying one stellar parameter.

The significant result to be noted here is that a much stronger
constraint on the varying parameter (in this case the chemical
composition) and the age can be obtained by invoking the \dtd\ than is
possible through the errorbox on the \hrd\ alone.  Of course, this is
not the definitive set of parameters for the best possible model, since
we have allowed only the chemical composition to vary, while fixing the
other parameters. We need to change all the parameters one by one
within the range allowed by the \hrd, and test for the remaining
parameter through the \dtd. But the strength of the \dtd\ lies in the
fact that starting from an initial guess model, we can get a very good
indication as to in which direction we should move in the
multi-dimensional parameter space to arrive at the best model. The task
of searching for the best possible combination of parameters is greatly
reduced by the \dtd\ through the additional information about the
location of the BCZ and HIZ. 

After several iterations, we were able to converge to a set of models
that satisfied the constraints on the \dtd\ for both \ttbcz\ and
\tthiz.  The final best models were selected on the basis of
comparisons of large separation values between the data and the models.
The absence of small separations in the data prevented us from
constraining the parameters further. We estimated errors on each
parameter to represent the range of values that would produce models
which are all consistent with the errorbox on the \dtd, as well as the
individual large separation values.  Needless to add, these models are
already consistent with the limits imposed on the \hrd. 
\myfigure{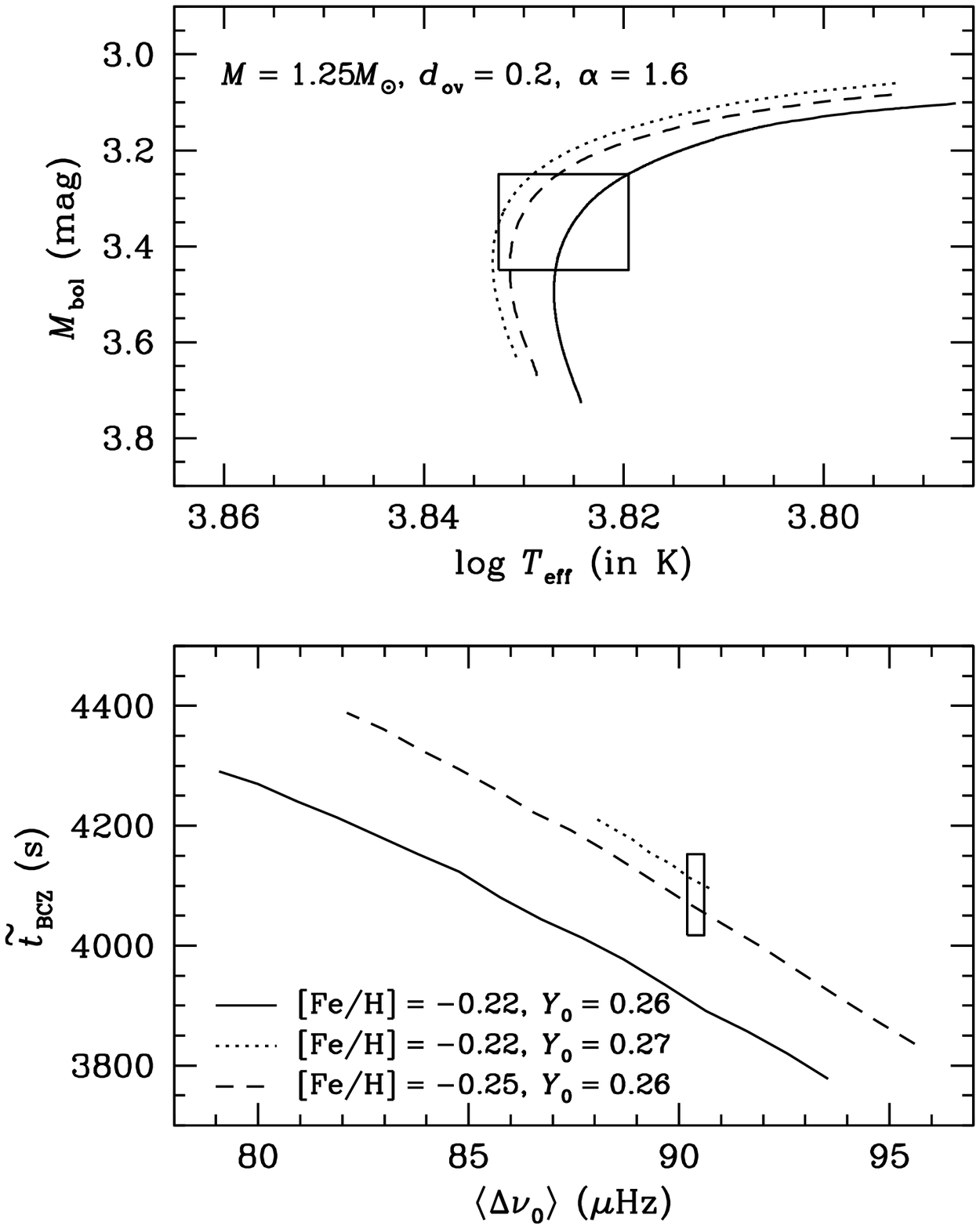}{{\it Top panel}: The position of
\object{HD49933} is shown on the \hrd. Three evolutionary tracks for
models with a given mass, overshoot and mixing length parameter, but
different chemical compositions are illustrated. {\it Bottom panel}:
The \dtd\ for \ttbcz\ is shown with the three tracks only for the
portion where they lie inside the box in the top panel. The derived
values of \dzz\ and \ttbcz\ for \object{HD49933} are represented by a
box.}{fig:hd49933_hrdt}{}

The parameters for our best models are shown in Table~\ref{tab:hd49933}
(stage~2), alongwith those used for the original input model. Clearly,
a remarkably good match with the input model was achieved. We should
add a caveat though -- as it turned out, both the input modelling and
the seismic analysis had utilised the same evolutionary code, CESAM;
this might have contributed partially to the exactness of the result.
However, the important point is not the goodness of the fit, but rather
the illustration of the use of a new seismic technique, the \dtd, which
led us to the final result. 
\begin{table}
\caption{The input parameters and results from a seismic analysis of
\object{HD49933} in a mock exercise are shown. The table is divided in
three parts, classified by stages of the seismic analysis: (0) initial
constraints provided to both modelling team and seismic interpretation
team; (1) results obtained from seismic analysis, prior to modelling;
(2) final results obtained from seismic modelling.
\label{tab:hd49933}
}
\begin{tabular}{ccccc}
\hline
\hline
Stage & Quantity        & Non-Seismic     & Seismic          & Original\\
  &                     & constraint      & value            & input   \\
\hline
0 & \teff               & $6700\pm 100$~K &                  & \\
  & $M_\mathrm{bol}$    & $3.35\pm 0.10$~mag &               & \\
  & \febyh              & $-0.32\pm 0.10$ &                  & \\
\hline
1 & \rhob/\rhobs        &                 & $0.465\pm 0.015$ & 0.467 \\
  & \ttbcz              &                 & $4085\pm 68$~s   & 4031~s \\
  & \tthiz              &                 & $4866\pm 49$~s   & 4805~s \\

\hline
2 & $M/M_\odot$         &                 & $1.24\pm 0.01$   & 1.25 \\
  & \xc                 &                 & $0.495\pm 0.010$ & 0.50 \\
  & $X_0$               &                 & $0.72\pm 0.01$   & 0.73 \\
  & $Z_0$               &      & $\phantom{0}0.01\pm 0.005$  & 0.01 \\
  & \dov                &                 & $0.20\pm 0.01$   & 0.20 \\
  & $\alpha$            &                 & $1.60\pm 0.05$   & 1.50 \\
\hline
\hline
\end{tabular}
\end{table}

\section{Conclusions}
\label{sec:concl}

We have investigated the possibility of using the \cdd\ for real
asteroseismic data by comparing the error in the separations propagated
from the observed frequencies against the sensitivity of the tracks
toward different stellar parameters.  The largest uncertainty in
calibrating the \cdd\ arises from the convective core overshoot
parameter and the chemical composition of the star.  However, the
effect of the mixing length parameter and the equation of state are
found to be relatively small. We estimate that if the chemical
composition is known independently, it would be possible to determine
the mass and age of a low mass star within $5\%$ using the average
frequency separations.  For a star with a convective core, the
uncertainty would increase due to the unknown extent of convective core
overshoot. 

We propose a new kind of seismic diagram (which we call the \dtd) for
low mass main sequence stars, connecting the acoustic location of sharp
features inside a star and its mean large separation. Since the
acoustic location of the base of the convective envelope or the
\ion{He}{ii} ionisation zone can be estimated to a precision of $\sim
95\%$, the position of a star on this diagram can be determined fairly
accurately.  The relation of such sharp features with the internal
stratification of the star and the dependence of the mean large
separation on its gross properties constitutes the physical basis
behind this diagram.  This diagram is designed to help in modelling the
star by exploiting the sensitivity of the acoustic location of the
sharp features toward different stellar parameters.

The use of the \dtd\ is illustrated through the seismic interpretation
of simulated data for a CoRoT primary target star, \object{HD49933},
carried out in the form of a hare and hound exercise. Firstly,
independent of comparison of actual models, the acoustic radii of the
base of the convective envelope and the \ion{He}{ii} ionisation zone
are recovered to a good accuracy by analysing the oscillatory signal in
the second differences of the frequencies. Further, all the stellar
parameters used in the original input model could be correctly
estimated through the seismic modelling based on the \dtd.

It is found that the \dtd\ can constrain the multi-dimensional space of
stellar parameters substantially and provide an iterative method of
quickly converging to the best model, starting from an initial guess.
The success of the method could be somewhat sensitive to the position
of the target star on the \hrd, since the effect of the various stellar
parameters on the \dtd\ is not uniform with respect to mass or age of
the star. We find that at least in the mass range of \mbox{$\sim
1$--$1.5\msun$}, this method is likely to be powerful. The efficiency
can be further enhanced if the \dtd\ can be used in conjunction with
the \cdd, which was unfortunately not possible in this particular hare
and hound exercise.

The application of asteroseismic diagrams to oscillation frequencies of
solar-type stars looks to be quite plausible, provided the dependence
of these diagrams on different unknown stellar parameters is taken into
account in the calibration process. On the other hand, this sensitivity
of seismic diagrams like the \dtd\ can actually be exploited to
construct reliable models for target stars based on the
model-independent results obtained from their oscillation frequencies.
Clearly, more tests are required with real data to refine these
methods.

\begin{acknowledgements} 
I would like to thank Ian Roxburgh, Caroline Barban, Thierry Toutain
and Eric Michel for their contribution to the results of the CoRoT Hare
\& Hound exercise presented here.  This work was supported partially by
CEFIPRA (Centre Franco-Indien pour la Promotion de la Recherche
Avancee) project No.\ 2504-3, and by the Research Fund, K.\ U.\ Leuven
under the grant GOA/2003/04. 
\end{acknowledgements}

\Online

\section{The classical C-D Diagram and its variations}
\label{sec:app}

We present a ``classical'' \cdd\ in Fig.~\ref{fig:d00d02}, involving
the large separations \dzz, and the small separations \dzt. The stellar
models used for this diagram were constructed using the OPAL equation
of state and had an initial chemical composition of $(X_0,Z_0) =
(0.70,0.02)$.  The mixing length parameter (in terms of the local
pressure scale height), $\alpha$, is $1.8$ and no convective core
overshoot was considered. 
\myfigure{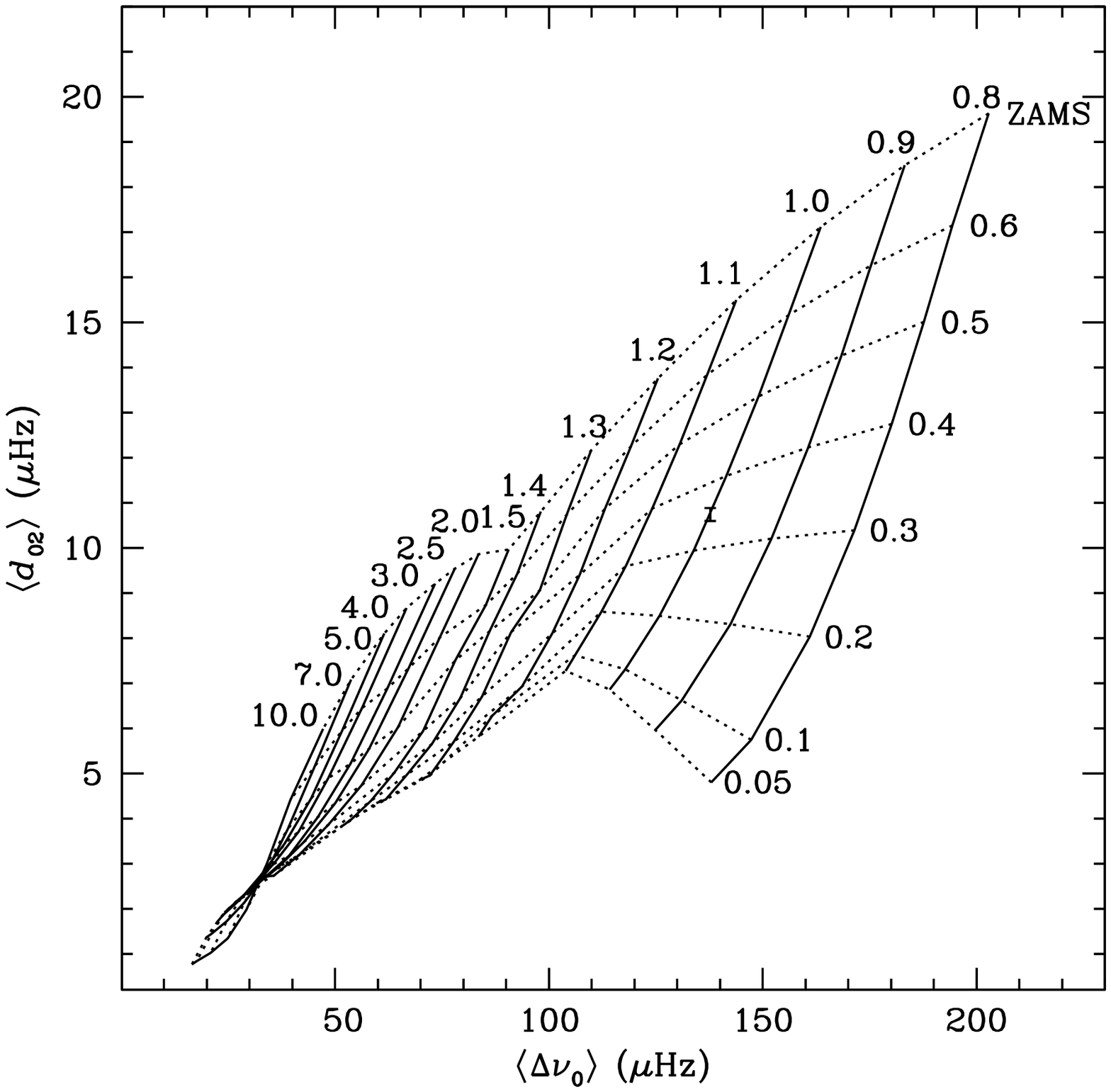}{The classical \cdd, plotting the average small
separation, \dzt, against the average large separation, \dzz, for a
grid of main sequence models between mass 0.8\msun\ and 10\msun.  The
bold lines follow the evolution of each stellar model with a given
mass, which is indicated at the top of the track. The dotted lines
connect the models at same stages of evolution, indicated by the
central hydrogen abundance, \xc.  Representative errorbar for \dzt\ of
a 1\msun\ star, corresponding to relative frequency errors of $10^{-4}$
is shown.}{fig:d00d02}{}

In Fig.~\ref{fig:d00r02}, the corresponding \cdd\ using the ratios of
separations is shown.  As pointed out by \citet{jcd04}, the filtering
out of the surface effects indeed increases the orthogonality of the
two effects regarding mass and age. Further, we find that this \cdd\
has improved ``resolution'' at higher masses, i.e., the lines
corresponding to masses greater than 2\msun\ do not collapse as much as
in the classical \cdd\ (cf.\ Fig.~\ref{fig:d00d02}). This implies that
the diagnostic capability of the \cdd\ can both be improved and
extended to higher masses if we use the ratios of separations instead
of the separations themselves.
\myfigure{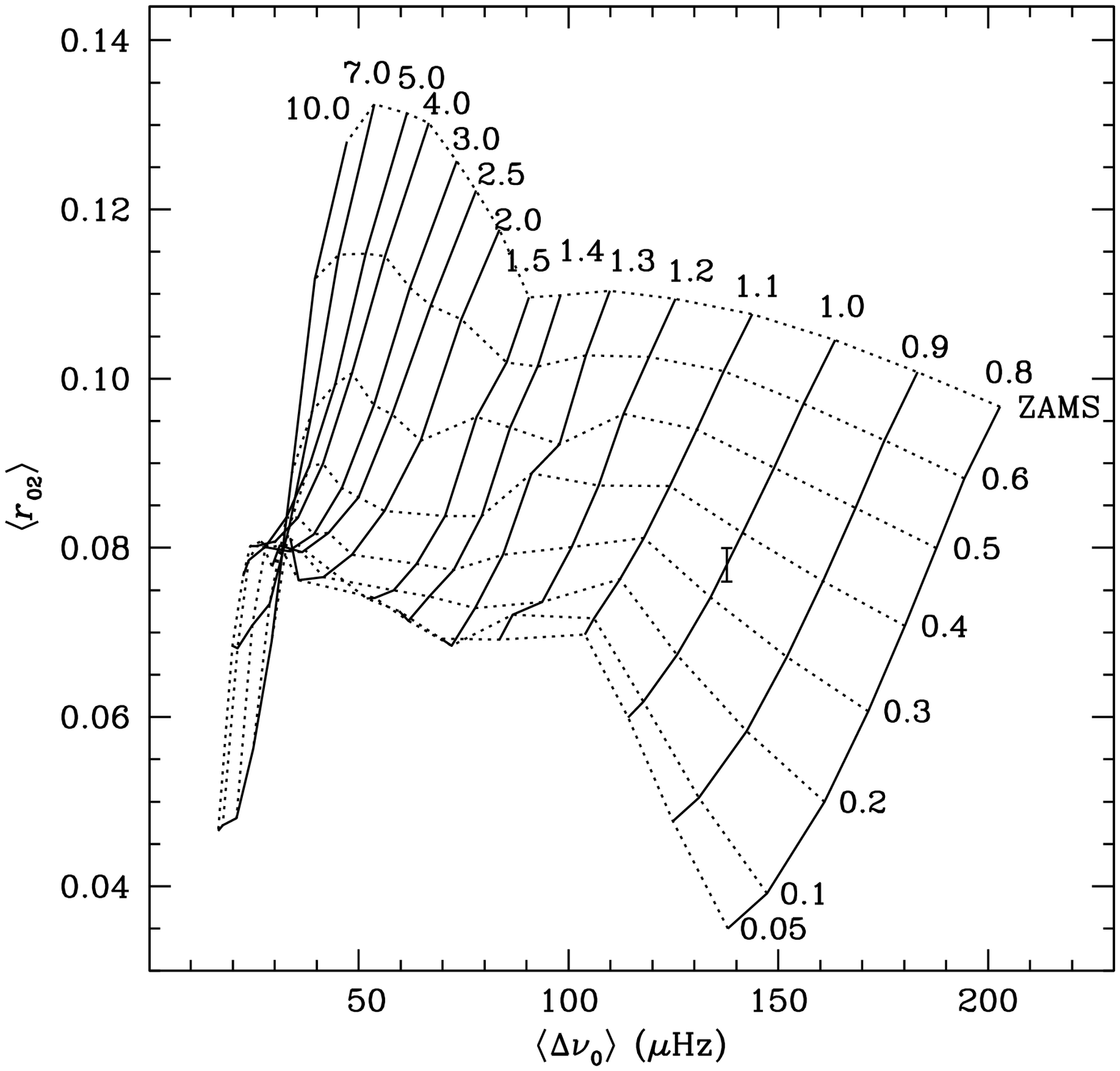}{A \cdd\ using the average ratio, \rzt, of the
small separation to the large separation. The stellar models used are
identical to those in Fig.~\ref{fig:d00d02}.}{fig:d00r02}{}

The kinks seen in the tracks for models with mass 1.2--1.5\msun\ are
not results of inconsistencies in the models, but are in fact related
to the evolution of the convective core. In this mass range, the
convective core grows in mass (or radius) for a significant part of the
main sequence evolution, before beginning to shrink later. The kink
corresponds to the evolutionary phase where this phase of growth of the
convective core ends and the shrinking phase begins.  This effect is
even more pronounced in models including convective core overshoot (see
Fig.~\ref{fig:dr02_ov}).

The construction of the \cdd\ illustrated above requires the
determination of a set of $\ell=0$ mode frequencies as well as $\ell=2$
frequencies of similar radial orders. In practice, however, we might
encounter the situation where only $\ell=0$ and $\ell=1$ modes are
identified for a target star, but not the $\ell=2$ modes. Following
Eqs.~(4) and~(5) of \citet{rv03}, we can define the 5-point separations
between these two sets of modes.  The \cdd s with these separations and
the corresponding ratios (Eq.~(8) of the same paper) are shown in
Fig.~\ref{fig:d00dd01} and Fig~\ref{fig:d00r01}.
\myfigure{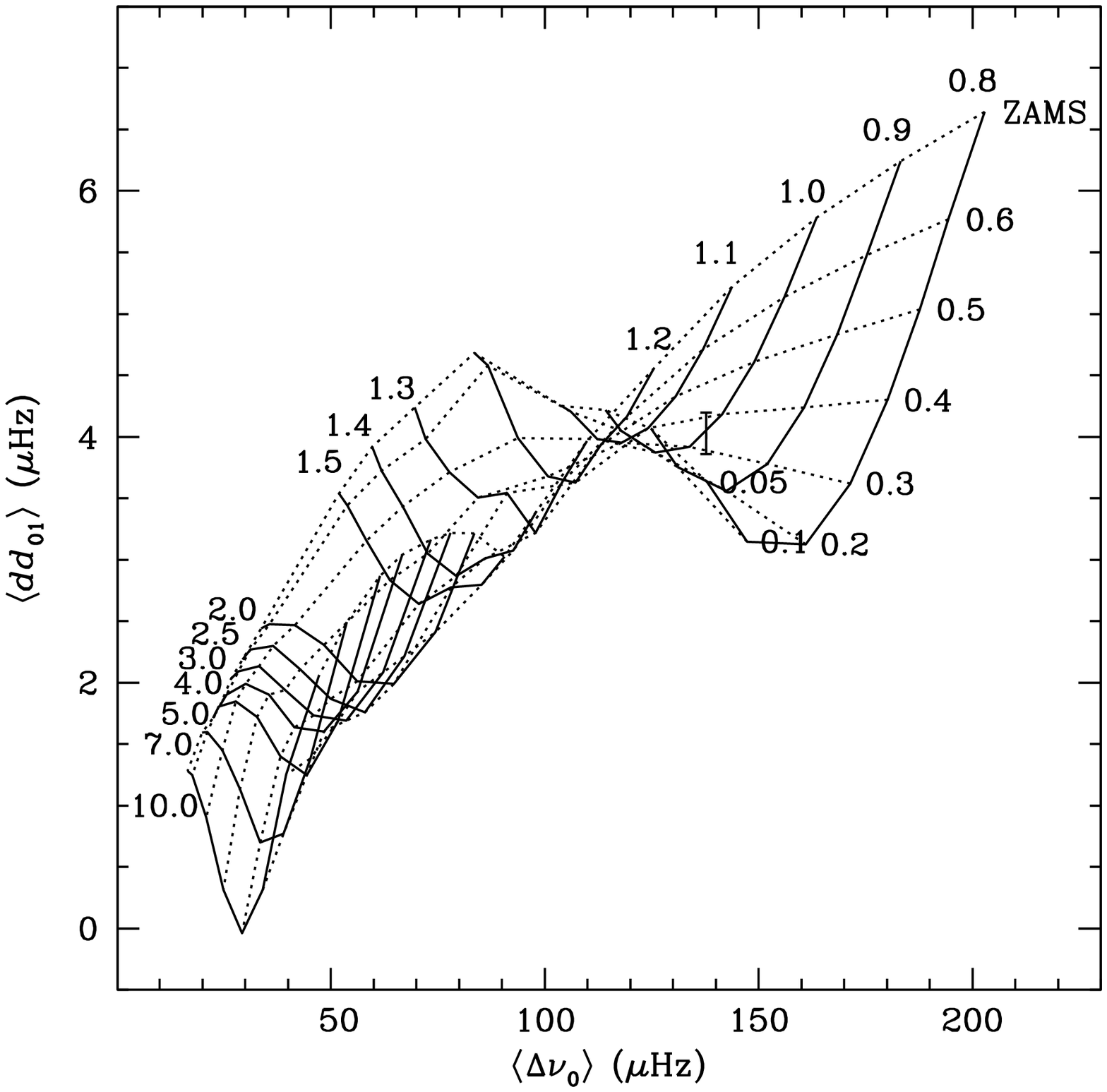}{An alternative \cdd\ using the average small
separations \ddzo. The stellar models used are identical to those in
Fig.~\ref{fig:d00d02}.}{fig:d00dd01}{}
\myfigure{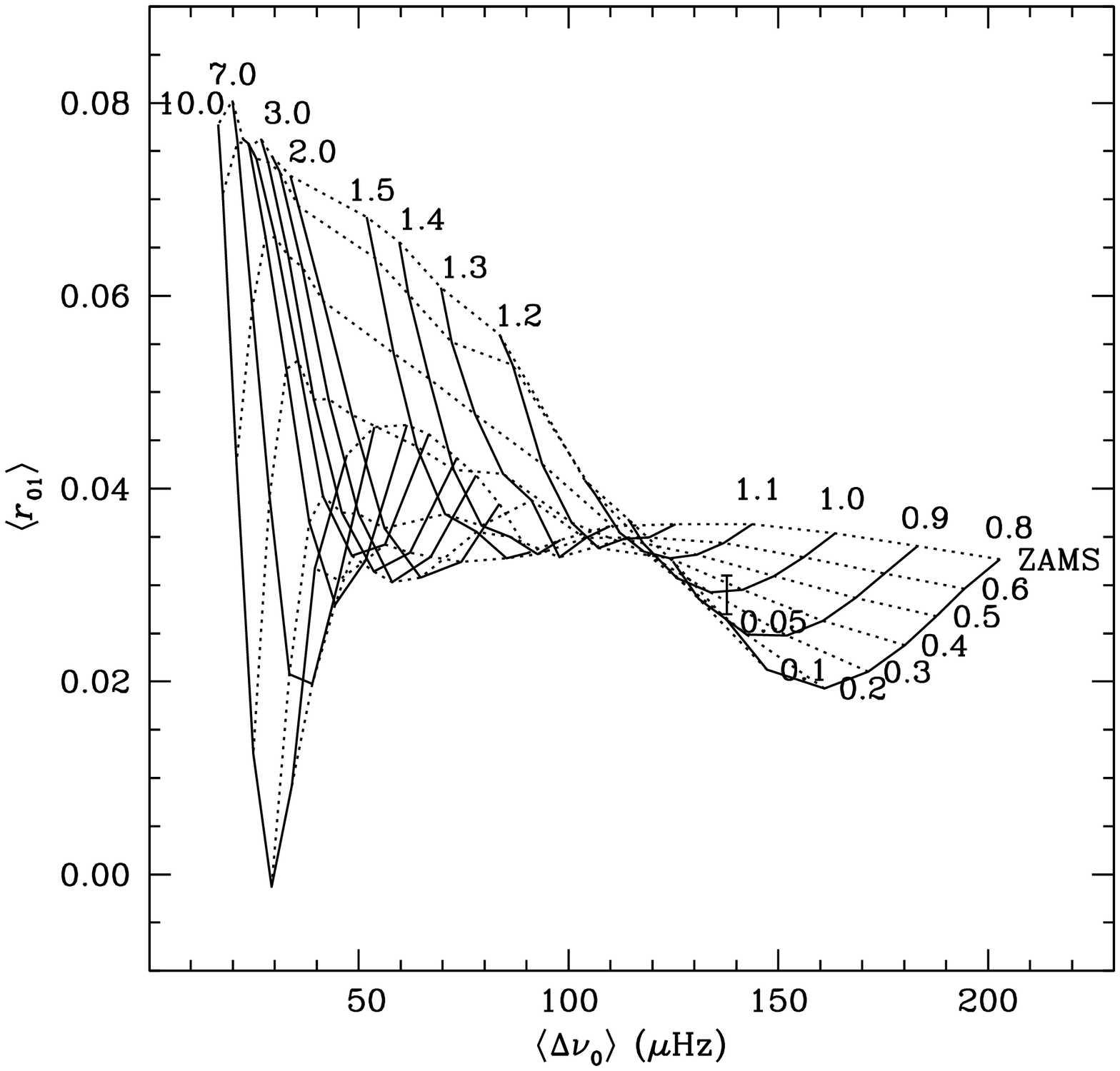}{A \cdd\ using the average ratio, \rzo, of the
small separation to the large separation. The stellar models used are
identical to those in Fig.~\ref{fig:d00d02}.}{fig:d00r01}{}

It is interesting to note the difference in the structure of the two
diagrams in Figs.~\ref{fig:d00d02} and~\ref{fig:d00dd01}. The \cdd\
with the \ddzo\ separations undergoes a ``twist'' at higher masses.
This is, in fact, to be expected from the behaviour of these
separations as functions of frequency. In Fig.~\ref{fig:nu_d02_dd01} we
plot both $d_{02}$ and $dd_{01}$ as functions of the scaled frequency,
$\nu/\sqrt{\rhob/\rhobs}$, where \rhob\ denotes the mean density, for a
sequence of models of mass 1.5\msun\ at different evolutionary stages
on the main sequence. The range of scaled frequencies used to compute
the average values in Figs.~\ref{fig:d00d02} and~\ref{fig:d00dd01} are
shown as well. We find that while the value of $d_{02}$ decreases
monotonically with age at all frequencies, at low frequencies the value
of $dd_{01}$ actually {\em increases} with age, flattening out to an
almost constant value at intermediate frequencies. At frequencies close
to the acoustic limit $dd_{01}$ decreases slightly with age. This
explains the ``twist'' in Fig.~\ref{fig:d00dd01}. Evidently, the range
of frequencies over which the average values are computed plays a
crucial role in determining the structure of the \cdd.  The physical
reason for this difference of behaviour between $d_{02}$ and $dd_{01}$,
as hinted by \citet{apc95}, possibly lies in the difference in the
position of the inner turning points of the $\ell=1$ and $\ell=2$ modes
close to the stellar core.  While the lowest frequency $\ell=1$ modes
would be reflecting off a layer close to the peak in the \bvf\ at the
boundary of the convective core, the $\ell=2$ modes may not be
penetrating that far.  The difference in behaviour of these two small
separations bears a clue to the evolution of the \bvf\ with age.
\myfigure{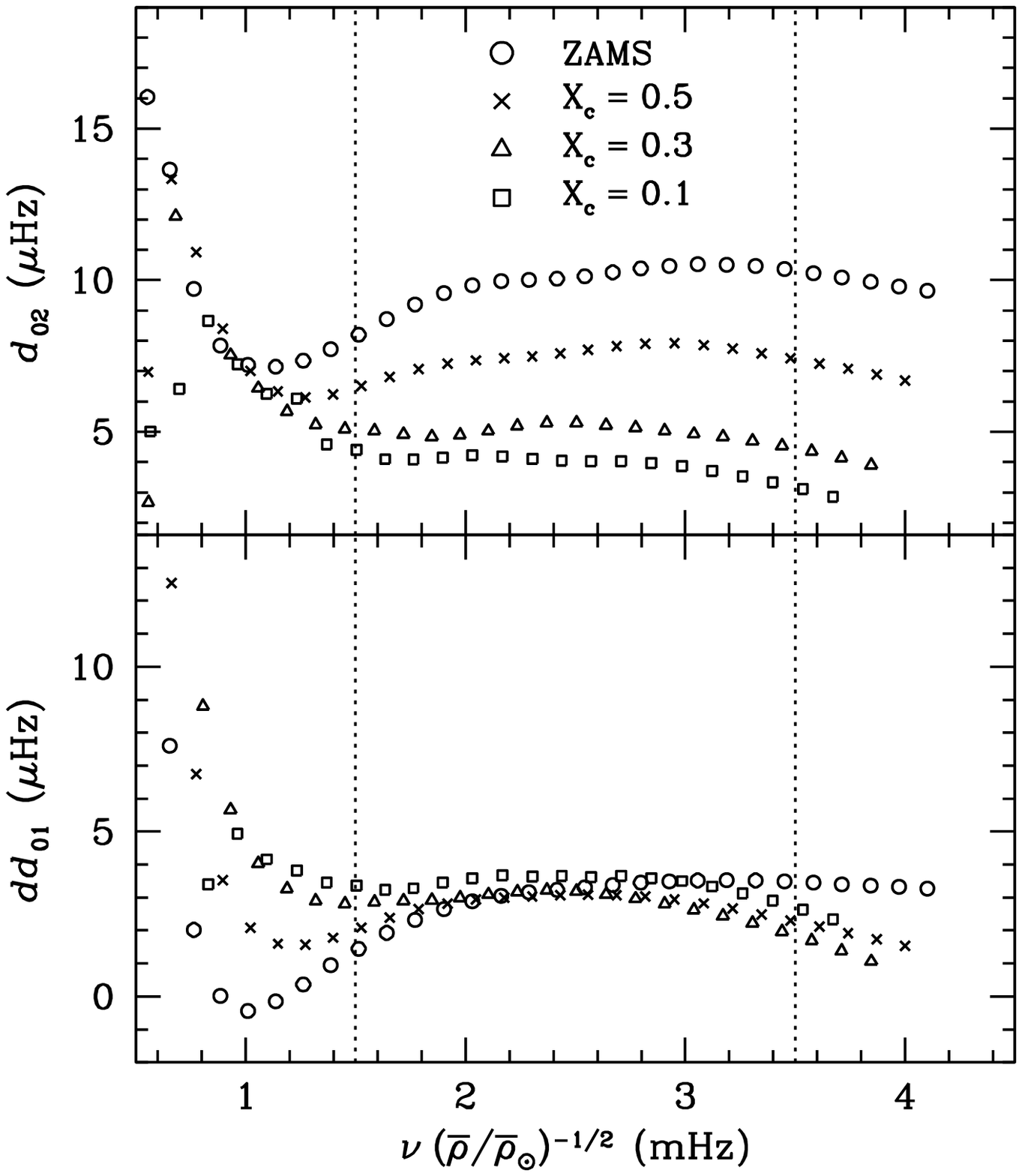} {The small separations $d_{02}$ ({\it top
panel}) and $dd_{01}$ ({\it bottom panel}) are shown as functions of
the scaled frequency for four 1.5\msun\ models at different stages of
evolution. The limits of scaled frequency used to compute the averages
for the \cdd s are indicated by the dotted lines.} {fig:nu_d02_dd01}{}

\end{document}